\documentstyle[epsf]{article}

\font\institute=cmss10

\title{Rigidity percolation  on aperiodic lattices}

\author{
A. Losev$^1$, F. Babalievski$^{1,2}$ \\
$^1${\institute Institute of General and Inorganic Chemistry}\\
{\institute Bulgarian Academy of Sciences, 1113 Sofia, Bulgaria}\\
$^2${\institute Institute for Computer Applications 1 (ICA1)}\\
    {\institute University of Stuttgart, 70569 Stuttgart, Germany}
  }

\date{October, 1997}

\begin{document}

\maketitle

\begin{abstract} 
We studied  the rigidity percolation (RP) model for
aperiodic (quasi-crystal)  lattices.  The RP thresholds (for
bond dilution) were obtained for several aperiodic lattices via
computer  simulation using the  ``pebble game'' algorithm. 
It was found that the (two rhombi) Penrose lattice is always floppy
in view of the RP model. The same was found for the Ammann's octagonal
tiling and the Socolar's dodecagonal tiling. In order to impose the 
percolation transition we used
so c. ``ferro'' modification of these aperiodic tilings. We studied as
well the  ``pinwheel'' tiling which has ``infinitely-fold'' orientational 
symmetry.  The obtained estimates for the modified Penrose, Ammann and
Socolar lattices are respectively: $p_{cP} =0.836\pm 0.002$, $p_{cA} =
0.769\pm0.002$, $p_{cS} = 0.938\pm0.001$. The bond RP threshold of the
pinwheel tiling was estimated to $p_c = 0.69\pm0.01$. 
It was found that these results are  very close to the Maxwell (the 
mean-field like) approximation for them.

\end{abstract}

   Modeling rigidity is a paradigmatic case of physical science as
it is classically conceived: the  consideration  of  an  elementary
mechanical model is used to  bring  some  light  in  an  altogether
different realm, for instance the behavior of matter at the atomic
scale. In
this way the questions why a construction such as the Eiffel  Tower
is stable or why glasses do  not  flow\cite{styklo}  are  linked  together.
 
 In  a  pioneering  work\cite{Maxwell}  Maxwell  sought to  know  when   a
mechanical construction of rigid bars  and  pin  joints  becomes  stable.  
The answer was: when  the  number  of independent constraints  
reaches  the number of degrees of freedom. But there is a next
task, which appeared much more difficult:
how one can  determine in a very large structure
which constraints are independent  and which are redundant.   

     Rigidity is an intuitively  clear  concept,  even  though  its
analysis soon reveals unusual aspects.  A  triangular  frame  formed
from three bars connected by pin joints is a rigid  body,  while  a
square is easily deformed. Regardless of the number  of  elementary
cells a construction of adjacent triangles  is  also  rigid   while
made out of squares it is still floppy. 
But in the latter case the
lack of rigidity may be thought as an effect of  the  finite  size:
if on a square lattice a periodic (helical) boundary conditions are imposed
it  would  be  a  rigid construction\cite{Thorp-priv} 
(see also \cite{9702249MD,Obukhov95}).
%%%% Of  course  speculations  about  the  properties  of
%%%%infinite objects are bound to be paradoxical but if  one  considers
%%%%the case of cyclic boundary conditions this dubious conclusion  may
%%%%be accepted. 

The next step in the analysis  lies  in  the procedure of 
``network  dilution''.  If randomly chosen  bars  are  removed
from the inside of a sufficiently large  rigid  structure  at  some
moment it loses this  integral property. 
Obviously
it can be carried out in the reverse direction: starting   with  an
unstable construction, bonds are added until it becomes rigid.
  (One
may note here that if it is carried out in an  orderly  fashion  it
allows to transform a generic square lattice  into  a  triangular  one --- or
vice  versa).  

A  more  general  approach  considers  an   arbitrary
collection of sites in space -- in the plane for instance -- which  are
joined to  their nearest neighbors, and to relate  the  change  in
behavior  with the numbers of possible bonds allowed, i.e. with the
coordination number. Indeed the bars and joints picture  
is a special case of the central-force percolation(CFP) model. In 
CFP  one can change the angles between bonds without cost of energy
and any motion which include change of bond lengths
would change the energy of the system. So one can differentiate
the CFP and bars and joints model 
(in this paper referred as  rigidity percolation). 
In the later case any changes in the bond lengths are not allowed and
the bond angles' changes are still  ``zero energy'' motions.
 
So the bars and joints picture could locate the place of the
rigid-to-floppy transition but could not give direct information
(e.g.) about the the elastic modulus critical behavior.
But this model picture has the huge computational 
advantage to make possible avoiding the forces equilibrium
calculations which usually scales with system size($L$) at criticality
as $L^{d +2}$ and faster ($d$ is the spatial dimension). 
This advantage  was not utilized for a long time since the numerical
simulations of this model remained the same as for general 
central--force
percolation --- via    forces equilibrium calculations.

A  recent
work by Thorpe \& Jacobs \cite{Thorp95} proposed an efficient way  
for overcoming  the
computational difficulties which arise in rigidity percolation
models. Instead of "perfect" lattices -- lattices which  bond lengths 
and bond angles are taken from a countable set -- their topological equivalents
were used: for such
%this??
 "generic" lattices, the connectivity is preserved but each 
bond and bond angle are taken
from continuous distribution. Moreover it is argued that the "perfect"
lattices are "atypical"  and  more natural are their generic counterparts.
Thorpe \& Jacobs \cite{Thorp95,Thorp96} also  turned the attention
to an efficient combinatorial algorithm\cite{pebl-alg} for constraints counting
called  the "pebble game" algorithm (see also \cite{DuxM95,Mouk96}.
 All that made possible estimating the central-force percolation
thresholds  without solving huge and badly conditioned sets of
linear equations.

In this work we present a computer simulation study of the rigidity
percolation in aperiodical (quasicrystalline)  twodimensional structures. 
We study the bond-dilution case of percolation on four aperiodic
lattices. Three of them are modification of  aperiodic 
lattices with "forbidden"
orientational symmetry: the two rhombi Penrose tiling 
(with five-fold symmetry), 
an octagonal tiling (known  as the Ammann's A$4$
tiling\cite{Ammann}) which is constructed by a square and a rhombus, 
and a dodecagonal tiling proposed by Socolar\cite{Socolar89}
constructed by a square, hexagon and a rhombus.

The  interest to such tilings came mainly after discovering of the
quasicrystals in 1984 \cite{Schechtman}. After the first observation of
icosahedral
quasicrystals,  soon after new metal alloys with one
periodic axis and 5(10)- 8- and 12-fold orientational symmetry (in the
perpendicular plane) were discovered.
 These four symmetries are likely the only ``non-crystallographic'' 
(rotational) symmetries which could be found in nature. 

 We modified the mentioned lattices by
adding bond through  these diagonals of the tiles, which are shorter
than the tile edge (See Fig.2; and in \cite{PhysA95}: Fig. $1${\bf c} 
and $1${\bf d}), the reason for that will be described below.  
The fourth aperiodic lattice we choose to study
was the so c. ``pinwheel'' tiling\cite{Senechal95,Aper97}: 
an aperiodic and deterministic tiling which edges are uniformly
distributed in all directions (Fig.1) --- in this sense -- a 
tiling with ``infinitely-fold'' orientational order. 

\begin{figure}
\centering
\leavevmode
\epsffile{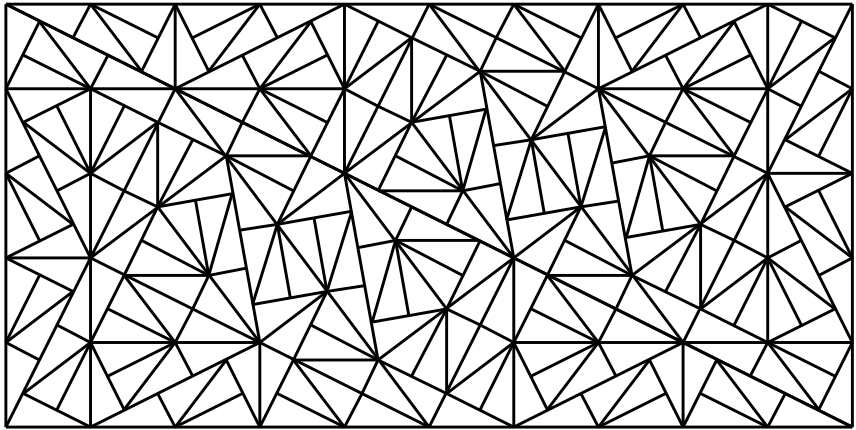}
\caption{
The  pinwheel tiling
}
\end{figure}

Indeed the orientational symmetry could not have direct relation with the
rigidity thresholds,
since the "pebble game" algorithm which we
use does not take in account the bond lengths and orientations.
Some indirect relation could be searched in the way the coordination
of neighboring sites is correlated. 
In this study we make comparison only with the mean coordination 
number.
The lattices we study here have coordination numbers between
$6$ and $4$ i.e. they can be ranged somewhere   between  the
paradigmatic cases of triangular and square  lattices.   

%================================================================

     A square lattice according  to  Maxwell's law  would be rigid
only if all bonds are present   ($p=1$)  and  of course there could not
be any redundant bonds in it.  The  (two rhombi) Penrose lattice,
the primer for deterministic   aperiodic   structure, has also a
coordination  number $ z=4$   and  failed  to  produce  any  clue  of
becoming rigid. 

So in order to see a rigidity transition one have to modify the
lattice in order to increase its mean coordination number.
The most natural modification is to put bonds between the lattice
sites if the distance between them is less than the tile edge length.
It was coined a name for this:  ferromagnetic modification, or, ferro
variant of an aperiodic tiling.

The 'ferro variant'
%(Fig.2) 
of the Penrose tiling has $z=4.76..$  and one could
expect that a rigidity percolation threshold should exist.

The non-modified variants of the octagonal and dodecagonal lattices
have mean  coordination numbers equal to $4$ and $3.63..$
respectively \cite{PhysA95}, so they have to be modified in an
analogous way. The ferro variants of these lattices includes new bonds
which are the short diagonals of the rhombuses in them.
As seen from the table the mean coordination $z$ is $5.17..$ and
$4.27..$ respectively.

The pinwheel tiling  consists  of identical triangles with  sides
in  the  ratio  $1:2:\sqrt{5}$ appearing in infinitely many orientations.
Inspection of the figure 1 shows that  in  about one  fifth  of   the
cases  two  of  the shortest sides   of   adjacent  triangles   are
co-linear  (forming   the  side  with  length  2  units  in  another
triangle) delimiting  thus  a  perimeter  with  4  points.
So this is a vortex-to-edge tiling. (See the M. Senechal's book in 
\cite{Senechal95})

The question here is how to deal with the vortices which lie on
a bond of another triangle. We choose  to think that   the
outer points of the pairs of such co-linear short   bonds  are  also
connected. Thus they  appear graphically as degenerated triangles of
zero area but in this type of study what matters is the
topology (connectivity) and not  the
geometry,(which  is  emphasized   in   the   concept   of   generic
network). The inclusion of  these  additional   bonds  has fixed  the
theoretical coordination number  to  $6$. In fact our largest sample of
the pinwheel tiling ($\approx 22 000$ sites) had a lower $z$. The ratio
between the whole  bonds and sites gave $z \approx 5.5..$ which 
probably is due to a larger bond deficiency at the borders of the sample.
For comparison, this pairs of values for the other lattices coincided
up to a less than  a percent.

\begin{figure}
\centering
\leavevmode
\epsffile{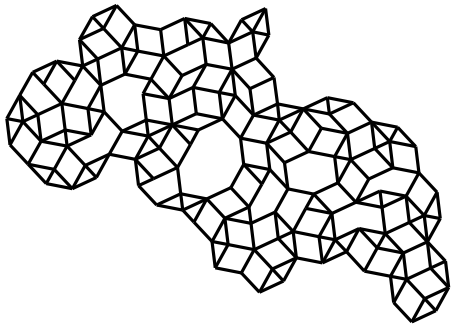}
\caption{ 
A part of rigid cluster (without the redundant bonds) in the
``ferro'' variant of the Penrose tiling ($p=0.83$).}
\end{figure}

%+ probably here some explanation how the theoretical values of $z$
%could be obtained.....
%====================================================================

We restrict our study to determining of the redundant bonds density
in  these lattices (for different bond dilutions).
As it will be shown further this is sufficient for estimation
of the rigidity percolation  threshold.
 
In general, a $d$-dimensional lattice with $n$ sites  and no bonds
between them will have $d.n - d(d-1)/2$ (in the plane $2 n -3$)
mechanical degrees of freedom
(or in the language of rigidity: floppy modes, or zero frequency
modes). If now bonds are put between sites the number of floppy modes
will decrease. If we neglect the angular forces, as it is accepted in
the central-force percolation model, each bond will decrease the floppy
modes at most by one. (Exactly said: by one or zero.)
 If no change occur in the number of floppy modes we speak about 
over-constraining or redundant bond.

Now, the task is one to differ, in a network of rigid bars and joints,
which bonds are redundant. In fact unambiguous decision for a certain 
bond could not exist for an already built construction. As was
mentioned 
previously, in a square with diagonals
one of the diagonals is redundant. In fact, each of the 6 bonds in
this construction could be thought as the redundant one. 
%Probably this is one of the reasons ..  so tough for analytical solving
 
In the count of floppy modes the case of redundant bonds  should  be
acknowledged  so $F=2n-(m-R)$, where  $F$ is the total number of floppy
modes for the given (twodimensional) lattice, $m$ is the number of
added
bonds and $R$ are the redundant amongst them.
Since the number of all bonds for a non-diluted (infinite) 
lattice is $z n / 2$ (where $z$ is the mean coordination number),
 the number of floppy modes per degree of freedom\footnote{of the 
unconstrained lattice} ($f = F/2n$)
can be written as:

$$
f = 1 - p\,\frac{zn}{2}\frac{1}{2n} - \frac{R}{2n}
$$

\noindent
where $p$ is the proportion of present bonds, or:

\begin{equation}
f = 1 - p\,\frac{z}{4} - r
\label{f-r}
\end{equation}

\noindent
where $r$ is the number of redundant bonds per degree of freedom.
If one neglects $r$, a mean-field-like (or Maxwell \cite{Maxwell}) prediction,
for the rigidity percolation threshold, could be done: $p_c^*= 4/z$.

The number of floppy modes($F$) is (roughly) proportional to the
number of  rigid clusters  for the system (if the isolated sites are
counted as well). Roughly, because a site may
belong to more than one rigid cluster. 
In analogy with the ordinary percolation
 model\cite{Stauffer} one can  argue \cite{Thorp95} that $f$
should behave as a free energy density, so its second derivative will
follow power law near to the (real) percolation threshold:

$$
f^{\prime \prime} \propto | p - p_c|^{-\alpha}
$$

\noindent
where $\alpha$ is a ``specific-heat like'' exponent. 
Integrating twice we can obtain the following form for $f$:

\begin{equation}
f(p) = b_1 +b_2 p + b_3|p -p_c|^{2-\alpha}
\label{f-form}
\end{equation}

Now comparing \ref{f-form} and \ref{f-r} we can use the data obtained for $r$
to estimate $p_c$ and (eventually) $\alpha$ (see Eq. \ref{r-form}
below).

   In  order  to  determine  the  rigidity percolation thresholds and 
the exponent $\alpha $ for the four aperiodic lattices the   following
procedure was established. The sites of a lattice  
are labeled with consecutive numbers and all their bonds are identified by the
$2$ numbers labeling the sites at their ends. The pairs of  integers
representing bonds are input with  some  probability  $p$  into  a
program which determines the number of  dependent  bonds  in  the
formed subset. Any such subset   describes  in  fact  a  particular
configuration.  The  collected  data  consists  in  the  number  of
dependent bonds monitored as a function of the varying  probability
$p$. We assume that they can be  approximated  satisfactorily (see
above) by  a function of the type

\begin{equation}
 r(p) = a_1 + a_2 p +a_3 | p -a_4 |^{a_5}
\label{r-form}
\end{equation}

\noindent
 where $a_1$, $a_2$, $a_3$ being arbitrary
parameters of no interest. While $a_4$ and $a_5$ should give
estimations for the percolation 
threshold ($p_c$) and the "specific-heat-like" critical exponent ($\alpha$) 
$a_4 \rightarrow p_c $ and $ a_5 \rightarrow 2 -\alpha $.

\begin{table}

\begin{tabular}{||l|c|c|c||}

\hline \hline
         & $\overline{z}$ & $p_c$ & $4/z$ \\
\hline
periodic triangular & $6$ &    $0.661 \pm 0.002\dagger$ & $2/3$ \\
\hline
pinwheel    & $6(5.5..)\ddagger$ & $0.69 \pm 0.01 $ & $2/3(0.727..)\ddagger$ \\
\hline
Penrose ("ferro") & $4.764...$ & $0.836 \pm 0.002 $ & $0.8396..$ \\
\hline
Octagonal(f)&       $5.17..$   & $0.769\pm 0.002$  & $0.774..$ \\
\hline
Dodecagonal(f) &    $4.27...$ & $0.938 \pm0.001$ & $0.937..$ \\
\hline \hline
%\end{minipage} 
\end{tabular}

$\dagger$
{\footnotesize  
a better result  
is given in \cite{Thorp95}: $0.6602\pm 0.0003$; }

$\ddagger$ {\footnotesize the second numbers are the actual values
  for 
the largest studied sample ($150 \times 150$)}\\
{\footnotesize (The size of the other lattices was up to $500 \times 500$)}

\caption{ Rigidity percolation thresholds (bond dilution) for
  aperiodic 
lattices. (The triangular lattice is studied to test the estimation method.)
}
\label{thresholds}

\end{table}

%===================================================================

The size of the lattices studied was of size up to  $\approx 500 \times
500$ tile edge lengths.
 The pinwheel tiling was smaller:
$150 \times 150$. It was generated by iterative applying\cite{Losev97}
the generating substitution rule\cite{Senechal95}. The other three lattice were
obtained
by a recursive implementation\cite{CPC90} of the de Bruijns'
 N-grid method\cite{deBruijn82}.

\begin{figure}
\centering
\leavevmode
 \epsffile{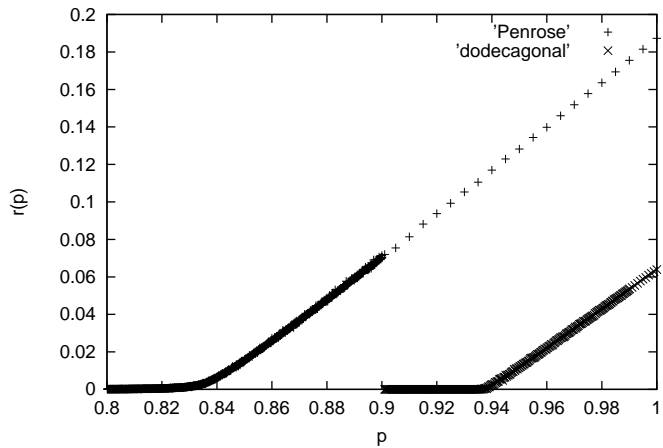}
 
\caption{
The redundant bonds per degree of freedom $r(p)$
  for the ``ferro'' variants of the Penrose lattice
(left curve) and the dodecagonal lattice.
Each data  point represents the result for {\em one} sample. 
The full lattices have $\approx 300\,000$ sites each.
}
\end{figure}

We counted the redundant bonds for lattices with different size and 
different bond dilutions, $q=1-p$ where $p$ is the probability
for present bond. We usually run {\em one sample} for each $p$ value,
but we used about 1000 different values of $p$ for each size. Usual
practice was to use different pseudo random number sequence for each
run.  

 Using a Marquard - Levenberg  based  optimization  routine
the parameters of a best fit with Eq.\ref{r-form} were estimated. 
This treatment of
data appeared robust regarding the parameter $a_4$ (the percolation
threshold). In opposite the other parameters appeared highly correlated
and were depending on system size and the interval of values for
$p$ within the ``measurements'' were made. The larger interval (say,
$ p \in [0.6 - 0.9]$ for the Penrose tiling) led to
smaller values for  $a_5$ (approaching $1$ as one could expect from the
figure).  One should suppose that using an interval
closer
to the threshold would give a better estimate, but this time the finite
size effects start to influence. We attempted finite-size scaling but
(probably due to the linear term in Eq.\ref{r-form}) we could not extract
consistent data.  It seems that
just counting the total number of redundant bonds is not sufficient to 
estimate the exponent $\alpha$. 
%As one can see from the figure this value
% would approach $1

As we already mentioned the estimates for $p_c$ were surprisingly
stable regarding the changes of systems size and the interval for $p$ 
used in  the fit. Of course some deviations were seen and we have
to made extrapolation to infinite size and to choose the interval for
varying of $p$. We used mostly the results for interval of $p$ within
6-7\% above and 2-3\% below the rough estimate for $p_c$.
 
 The obtained results are  summarized  in  Table \ref{thresholds}. For   a
triangular lattice  the  percolation  threshold  has  been  already
established with great accuracy \cite{Thorp95} to be $ 0.6602\pm
0.0003$  while  the  Maxwell prediction is  $2/3$ The  procedure utilized here
gives  $0.661 \pm 0.002$ which  supports our results for the other
lattices.
%suggests  that despite  its  roughness
%leads  to  valid results.
% if we were able to extend the curves beyond $p=1$.

We check these results by adding a kind of bus-bars to two opposite edges
of the lattice sample\cite{DuxM95}. We simply used $p=1$ when entering into the
pebble game program the bonds within the left and right edge of the
``sample''. After reading all bonds we added {\em one more} bond to connect
a site from the left edge to a site to the right one. If a spanning rigid
cluster has already existed between these edges, the new (long-range) 
bond should be redundant.
 We studied in this way the largest lattice sizes by fixing three values
for $p$: $p=p_c-\Delta p_c; \; p=p_c; \; p=p_c + \Delta p_c$; where
$\Delta p_c$ was equal to the estimated ``error bars'' given in the
Table 1.
 We made typically 100 runs for each value of $p$.
It appeared that we have chosen the proper interval for $p$  to estimate
$p_c$ in our fits of $r(p)$. 
%We usually detected less than 10 spanning
%samples be

When compare the entries in the last two columns of the table one can
see that the mean-field like approximation works very well and it 
becomes better when the percolation threshold is closer
to $1$. When look on the curves on Fig.3 one can mention (in fact
H.J. Hermann was who mentioned) that almost 90\% of the bonds
added above the percolation threshold  are redundant. So, the building
parts of the spanning rigid cluster exist even below the threshold
and only few bonds are needed to connect them in the rigid structure which
spans the sample.

\smallskip

{\em In conclusion} one may summarize the results of this work as
follows:
it was studied for the first time the rigidity percolation model for  
some aperiodic lattices. Four typical representatives of these
lattices were studied: the Penrose tiling from two rhombuses, the
Ammans'
octagonal tiling,  the Socolars' dodecagonal 
tiling, and the ``pinwheel'' tiling constructed by J. Conway.
It was shown that the counting of redundant bonds in 
rigidity percolation models on these tilings is sufficient to locate
 the percolation threshold with a good precision.
The rigidity percolation ``generic'' thresholds for bond dilution were
estimated and compared with the Maxwell approximation. 
The results show that the critical
region is very narrow for this lattices as is the case for
triangular lattice, so the Maxwell approximation (to neglect
the redundant bonds) gives very good estimates for the percolation thresholds.

It would be interesting, the obtained here ``generic'' thresholds 
to be compared with
results from force equilibrium calculations on ``perfect'' aperiodic
lattices. One could expect that the difference should be smaller than
for triangular lattices, since so c. diode effect in aperiodic lattices is
less pronounced.

\smallskip

We acknowledge the discussions with M. Thorpe as well sending us the
Jacobs-and-Thorpe ``pebble game'' program.
One of us (F.B.) acknowledge the support from the German Academic
Exchange
Foundation (DAAD). F.B. also thanks to V. R\"ais\"anen and H.J.
Herrmann for the helpful discussions and to ICA1 for the hospitality.


\begin{thebibliography}{99}

\bibitem{styklo} P. G. Wolynes, Nature, {\bf 382} (1996) 495.

\bibitem{Maxwell} J. C. Maxwell, Phil. Mag. ser 4 {\bf 27} (1864) 294;
contemporary edition of the manuscript could be found in:
The Scientific Letters and Papers of James Clark Maxwell, (edited by
P.M. Harman) v. {\bf II}, Cambridge Univ. Press, Cambridge 1995.

\bibitem{Thorp-priv} M. F. Thorpe - private communication.

\bibitem{9702249MD} C. Moukarzel, P. M. Duxbury, {\tt lanl} e-print:
{\tt cond-mat/9702249}

\bibitem{Obukhov95} S. P. Obukhov, Phys. Rev. Lett. {\bf 74} (1995) 4472. 

\bibitem{Thorp95}
D. J. Jacobs, M. F. Thorpe, Phys. Rev. Lett. {\bf 75} (1995) 4051.

\bibitem{Thorp96}
D. J. Jacobs, M. F. Thorpe, Phys. Rev. E {\bf 53} (1996) 3682.

\bibitem{pebl-alg}
B. Hendrickson, SIAM J. Comput. {\bf 21} (1992) 65.

\bibitem{DuxM95} C. Moukarzel, P. M. Duxbury, Phys. Rev. Lett. {\bf 75}
  (1995) 4055.

\bibitem{Mouk96} C. Moukarzel, J. Phys A {\bf 29} (1996) 8079.



\bibitem{Ammann}
R. Amman, B. Gr\"unbaum, G. C. Shephard, Discrete \& Computational
Geometry {\bf 8} (1992) 1-25.

\bibitem{Socolar89} J. E. S. Socolar, Phys.Rev B {\bf 39} (1989) 10519.

\bibitem{Schechtman} D. Shechtman et all., Phys. Rev. Lett. {\bf 51} 
(1984) 1951.

\bibitem{PhysA95} F. Babalievski, Physica A {\bf 220} (1995) 245.

\bibitem{Senechal95} J. H. Conway (unpublished) --- see Ch.7 in:
M.~Senechal, {\em Quasicrystals and Geometry}, Cambridge Univ. Press,
Cambridge 1995.

 \bibitem{Aper97} C. Radin p.499 in: {\em The Mathematics of  
Long-Range Aperiodic Order} (ed.: R. V. Moody) NATO ASI Series C {\bf 489},
Kluwer Academic Publishers, Dodrecht/Boston/London 1997.



\bibitem{Stauffer} D. Stauffer, A. Aharony, {\em Introduction to
Percolation Theory}, Taylor and Francis, London 1992, 1994.

\bibitem{Losev97} A. Losev (unpublished).

\bibitem{CPC90} F. Babalievski, O. Peshev, Comp. Phys. Commun. {\bf 60} (1990)
  27.

\bibitem{deBruijn82} N. G. de Bruijn, K. Ned. Akad. Wetensch. Proc. 
{\bf A84} (1981) 51.

\end{thebibliography}
\end{document}